\documentclass[12pt,a4paper]{article}

\usepackage[dvips]{graphicx}

\begin{document}
\thispagestyle{empty}

\mbox{}
\vspace{1cm}
\begin{center}
{\Large{\bf The impact of non-linear functional responses on the}}

\smallskip

{\Large{\bf long-term evolution of food web structure}}

\vspace{0.5cm}

{\it Barbara Drossel$^{\rm a}$, Alan J. McKane$^{\rm b}$, Christopher 
Quince$^{{\rm c}\,\dag}$} 
\\
\bigskip
$^{\rm\, a}$Institut f\"ur Festk\"orperphysik, Technische Universit\"at 
Darmstadt, Hochschulstr.~6, 64289 Darmstadt, Germany\\
$^{\rm\, b}$Department of Theoretical Physics, University of Manchester, \\
Manchester M13 9PL, UK \\
$^{\rm\, c}$Department of Physics and Astronomy, Arizona State University, \\
Tempe, AZ 85287-1504, USA \\
\end{center}

\vspace{2cm}

\begin{abstract}
We investigate the long-term web structure emerging in evolutionary food web 
models when different types of functional responses are used. We find that 
large and complex webs with several trophic layers arise only if the 
population dynamics is such that it allows predators to focus on their best 
prey species. This can be achieved using modified Lotka-Volterra or 
Holling/Beddington functional responses with effective couplings that depend 
on the predator's efficiency at exploiting the prey, or a ratio-dependent 
functional response with adaptive foraging. In contrast, if standard 
Lotka-Volterra or Holling/Beddington functional responses are used, long-term 
evolution generates webs with almost all species being basal, and with 
additionally many links between these species. Interestingly, in all cases 
studied, a large proportion of weak links result naturally from the evolution 
of the food webs.
\end{abstract}

\vspace{2cm}


\vspace{2cm}

$^{\dag}\,$The authors are listed in alphabetical order.

\newpage

\section{Introduction}
\label{intro}
The debate over what stabilises complex food webs has been an active
one for over a quarter of a century (McCann, 2000).  While it is found
that the stability of a randomly linked food web model tends to decrease 
with the proportion of links and the number of species (May, 1972), real 
food webs display a high degree of stability, in spite of being very 
complex. Clearly, real food webs have features that are ignored in randomly 
linked models. For example, the distribution and strength
of links in real food webs are far from random. More realistic
approaches use link values taken from real food webs (Yodzis, 1981),
or assemble a food web by repeated addition of species from a large
species pool that contains different types of species such as
``plants'', ``herbivores'', ``carnivores'', and ``top
predators''(Morton \& Law, 1997; Law, 1999). Using Lotka-Volterra type
dynamics, assembly models of this type lead to large complex
webs. These Lotka-Volterra type models lack nonlinear effects such as predator
saturation, which many consider important for food web dynamics. The 
species pool is usually composed of a large number of species which are 
generated in an ad hoc manner, by pre-defining the possible types and 
value ranges of links. A complementary approach to food web stability 
focuses on more realistic dynamics (that include predator saturation, for
instance) of small systems consisting only of a few species. 
These small systems can be viewed as a small part of a food web. It has 
been found that the presence of many weak links and the use of nonlinear 
growth rates, that prevent predators from feeding successfully on many prey 
at the same time, stabilise the dynamics of these small systems by reducing 
population fluctuations (McCann {\it et al.}, 1998). Again, the possible 
types and values of links are pre-defined in these studies.

In real food webs, the link strengths and the linkage pattern can
change with time through invasion and replacement with related
species, and through adaptive evolutionary processes (Thompson,
1998). It is therefore desirable to investigate food web models that
implement such long-term changes due to modification of existing link
strengths and linkage patterns. In the following, we call this
long-term change of the food web ``evolution'', with the meaning of
``development of the food web structure in time'', while the ``true''
biological evolution is of course happening on the level of
individuals.  Our approach is different from the species assembly
models insofar that the new species are not picked irrespective of the
web composition, but are similar to existing species.

Considering the long-term change of food webs adds a new dimension to
the complexity-stability debate: even though species die out and are
replaced with others during the course of time, ecosystems as a whole
persist in time. Except for rare catastrophic events due to external
causes, food webs preserve their general structure. This means that
typical structural features of food webs, such as the total number of
species or the number of trophic layers, do not collapse to zero, but
rather fluctuate around some mean value far away from zero. This
definition of stability is perhaps more relevant for understanding the
complexity of ecosystems, than some of the other definitions found in
the literature. Persistence in time of a complex ecosystem of course
also requires a certain kind of stability of the population dynamics:
while population dynamics does not need to reach fixed points,
oscillations must be small enough such that the population sizes of
the majority of species remain positive after the introduction of a
mutant or invading individual. If this were not so, each change in the 
web composition would lead to a collapse of the web, and complex
ecosystems could never arise. 

In this paper, we therefore study evolutionary food web models and
investigate under which conditions these models show a persistent
complex structure with several trophic layers. The evolutionary model
that we use was studied previously with specific types of population
dynamics (Caldarelli {\it et al.}, 1998; Drossel {\it et al.}, 2001). 
Here, we explore the effect of changing the functional response used 
in the model. We chose a representative selection of
functional responses found in the literature and modified them such
that they are suitable for many-species communities.  Our aim here is
to show that biologically realistic requirements, in particular the
capability of predators to adjust their feeding rates and focus on
their best prey, are essential to the construction of large stable
complex webs. 

\section{The model}

The governing equation for the population $N_{i}(t)$ of species $i$ is taken
to have the form
\begin{eqnarray}
\frac{dN_{i}(t)}{dt} &= &\lambda\sum_{j}N_{i}(t)g_{ij}(t) - 
\sum_{j} N_{j}(t)g_{ji}(t) \nonumber \\ && - d_{i}N_{i}(t)
 -\sum_j \alpha_{ij}N_i(t)N_j(t)\,,  
\label{balance}
\end{eqnarray}
which naturally accounts for the four processes which lead to a change
in the population number $N_{i}(t)$. The function $g_{ij} (t)$ is the
functional response, i.e.~the rate at which an individual of species
$i$ feeds on species $j$; it depends on the population sizes, and its
analytical form will be specified below.  The constant $\lambda$ is
the ecological efficiency at which consumed prey are converted into
predator offspring.  The first term describes population growth due to
food consumption, while the second term describes population decline
due to predation. The death rate, $d_{i}$, will be assumed to be equal
to 1 for all species. The last term describes direct interference
competition between two predators $i$ and $j$. It is needed to
represent intra-specific interference competition, which ensures
bounded solutions when the functional response is independent of
predator density. This term also allows the incorporation of
inter-specific interference competition into such models, which
facilitates comparison with the predator dependent functional
responses.

On a larger time scale, species undergo changes, either by replacement
with invading similar species, or by evolutionary change. In order to
implement these changes, each species was characterised by a set of 10
out of 500 possible features (Caldarelli {\it et al.}, 1998). This
representation gives a measure of similarity between species (the
number of features they have in common) and allows for ``mutations''
or ``invasions'' by randomly replacing one feature of one individual
with another. Scores between features were assigned in a random and
asymmetric fashion. They are a measure of how useful a feature is for
its carrier at feeding on a species carrying another feature.  The
interactions, or scores, $a_{ij}$, between species were obtained by
adding the mutual scores between all pairs of features carried by the
two species.  Positive scores indicate that the first species can feed
on the second species and negative scores mean that the first species
is consumed by the second. In the latter case the score was set to zero, 
since this information was already contained in the (positive) score where 
the two species are interchanged. Therefore all scores were such that 
$a_{ij} \geq 0$. The external resources were represented as an additional 
species of fixed (and large) population size, which does not feed on any 
species. The evolutionary dynamics consisted of the following steps: 
starting from some initial species configuration (usually the external 
resources plus one basal species feeding on it), one individual was 
picked at random, and one of its features was changed. Under the dynamics 
(\ref{balance}), this new species either died out, or added to the system, 
or drove one or several other species to extinction. When the dynamics 
had reached an equilibrium, the next ``mutation'' occurred, and the 
process was repeated.

In the following, we list the different functional responses used in our 
computer simulations. Lotka-Volterra dynamics are recovered by taking 
\begin{equation}
g_{ij}= a_{ij}N_j\, .
\label{LV}
\end{equation}

A  Holling type II functional response, which implies saturation of
consumption rates at high prey abundance, is given by
\begin{equation}
g_{ij} = \frac{a_{ij} N_j}{1+\sum_k b_{ik}N_k}\,,
\label{generalholling}
\end{equation}
where the sum in the denominator is taken over all prey $k$ of species
$i$. 

More complicated functional responses for multi-species systems can
only be found in the recent literature.  Arditi and Michalski (1996)
suggest the following generalised Beddington form:
\begin{equation}
g_{ij} = \frac{a_{ij} N_j}{1+\sum_k b_{ik}N_k+\sum_lc_{il}N_l}\,,
\label{generalbeddington}
\end{equation}
where the first sum is again taken over all prey $k$ of species $i$,
and the second sum is taken over all those predator species $l$ that
share a prey with $i$. We chose the $c_{il}$ such that individuals
belonging to the same species competed more strongly with each other
than individuals belonging to different species.

Arditi and Michalski (1996) also suggest the following ratio-dependent
functional response, which implements the idea that predators share
the prey:
\begin{equation}
g_{ij} = \frac{a_{ij} N_j^{r(i)}}{N_i+\sum_{k\in R(i)}b_{ik}N_k^{r(i)}}\,,
\label{generalratio}
\end{equation}
with the self-consistent conditions 
$$N_j^{r(i)}=\frac{\beta_{ji}N_i^{C(j)}N_j}{\sum_{k\in
C(j)}\beta_{jk}N_k^{C(j)}},\quad
N_k^{C(j)}=\frac{h_{jk}N_j^{r(k)}N_k}{\sum_{l\in R(k)}h_{lk}N_l^{r(k)}}\,
.$$
Here $\beta_{ij}$ is the efficiency of  predator $i$ at consuming
species $j$, $h_{ij}$ is the relative preference of predator
$i$ for prey $j$, $R(i)$ are the prey species for predator $i$,
$C(i)$ are the species predating on prey $i$, $N_j^{r(i)}$ is the part
of species $j$ that is currently being accessed as resource by species
$i$ and $N_k^{C(j)}$ is the part of species $k$ that is currently
acting as consumer of species $j$.

Finally, in a recent paper (Drossel {\it et al.}, 2001), we used the 
ratio-dependent expression
\begin{equation}
g_{ij}(t) = \frac{a_{ij}f_{ij}(t)N_j(t)}{bN_j(t)
+\sum_k \alpha_{ki}a_{kj}f_{kj}(t)N_k(t)}\,,
\label{ourgij}
\end{equation}
where $f_{ij}$ is the fraction of its effort (or available searching time)
that species $i$ puts into preying on species $j$. These efforts are
determined self--consistently from the condition
\begin{equation}
f_{ij}(t) = \frac{g_{ij}(t)}{\sum_k g_{ik}(t)}.\label{eff}
\end{equation}
This condition is such that no individual can increase its energy
intake by putting more effort into a different prey. The sum in 
(\ref{ourgij}) is over all species $k$ which are predators of $j$.
The competition strength $\alpha_{ik}$ is set equal to one only for 
$i=k$, and is smaller than 1 otherwise. For this model, or for any 
other model with a functional response which depends on predator 
density, the last term in Eq.(\ref{balance}) should be dropped. 

\section{Simulation results: food web structure}

\begin{figure}
\begin{center}
\includegraphics*[width=5cm]{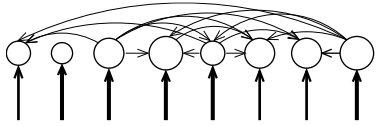}
\caption{Example of a food web resulting from an evolutionary model
with Lotka-Volterra population dynamics. The arrow direction indicates
the flow of resources, and the arrow thickness is a measure of link
strength. Links are only drawn if a species obtains more than 1
percent of its food through that link. The radius of the circles
increase logarithmically with population size. \label{fig1}}
\end{center}
\end{figure}

Let us first present results for the simplest type of population
dynamics, which is Lotka-Volterra dynamics, Eq.~(\ref{LV}). Starting
with one species and the external resources, input at a rate $R$, we 
allowed for enough time for the system to evolve towards its characteristic 
stationary structure. A typical food web resulting after sufficiently 
long time is shown in Fig.~\ref{fig1}. All species feed on the external
resources, and additionally on other species.  Only for short
transient periods of time  does the web have species on the second
trophic level, and virtually never on the third. We also performed
simulations that started from a complex web with several trophic
layers, which was stable under the population dynamics. However, the
evolutionary dynamics caused this structure to collapse, and the
resulting structure was again similar to that shown in Fig.~\ref{fig1}. 
These results were obtained without inter-specific direct 
competition, i.e. with $\alpha_{ij}=c\delta_{ij}$ where $\delta_{ij}$
is the Kronecker delta defined by $\delta_{ij} = 0$ for $i \neq j$
and $\delta_{ii} = 1$. They are still true when direct inter-specific 
competition is included, as in a different evolutionary model described 
by L\"assig {\it et al} (2001) and also in our model if we set  
$\alpha_{ij} = c\rho_{ij}(1 + q_{ij})/2$ in Eq.~(\ref{balance}). Here 
$\rho_{ij}$ is equal to one if $i$ and $j$ share at least one prey and 
is zero otherwise, and $q_{ij}$ is the fraction of discrete features 
shared by the two species. 


Our interpretation of these results is as follows. When a modified
species that can feed on more prey than a parent species is introduced
into the model, the modified species takes over. This is due to the
linear growth rates of Lotka-Volterra systems that allow a species to
feed on each of several prey just as easily as it would feed
on one prey, if only this prey was present. If allowed to evolve, an
artificially constructed Lotka-Volterra food web will therefore tend
to a simple web structure where each species feeds on many other
species and in particular on the most abundant resource, the external
environment. Clearly, this feature of the model is unrealistic as
there are evolutionary trade-offs that do not allow species to feed
very easily on an arbitrarily large number of species. In particular,
species cannot normally feed on external resources and on many other 
types of prey. 

In the light of these arguments, a Holling or Beddington type
functional response is more realistic, since it naturally limits the
total amount of prey taken by a predator. Indeed, the need for 
nonlinear growth rates in the context of multispecies communities 
has often been discussed in the literature (Pimm, 1991; Rosenzweig, 1995; 
Vandemeer {\it et al.}, 2002). Functional responses of the form
(\ref{generalholling}) were used by McCann
\emph{et al} (1998) in their investigation of the effects of weak
links. They have the biologically meaningful feature that predators
cannot maintain a high feeding efficiency on many prey at the same
time, and they have a stabilising effect on the dynamical equations,
at least for subsystems consisting of up to four species (McCann {\it et al.}, 
1998). Additional direct inter-specific competition can be introduced in a
natural way by using functional responses of the Beddington form
(\ref{generalbeddington}). However, when evolutionary dynamics by
modification of existing species are added to either of these models,
we find again that complex webs with several trophic layers are not
stable and that after some time the webs consist merely of basal
species, with a structure very similar to that shown in Fig.~\ref{fig1}. 

\begin{figure}[t]
\begin{center}
\includegraphics*[width=6cm]{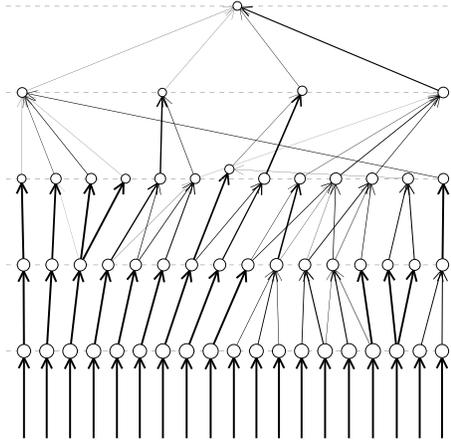}
\caption{Example of a food web resulting from an evolutionary model with 
Lotka-Volterra population dynamics and adjusted interactions ($\delta=0.2$). 
The same conventions apply as for Fig.~\ref{fig1} except that vertical 
position is now proportional to the average path length from the species 
to the environment weighted by diet proportions. Notice that the population
sizes vary much less across trophic levels than when a ratio-dependent 
functional response is used --- as in Fig.~\ref{fig3}, for instance.
\label{fig2}}
\end{center}
\end{figure}

It appears that even for the Holling or Beddington type functional
response too many species can feed on the same food source, and that
the model does not have a mechanism that prevents possible links (i.e. 
$a_{ij}>0$) from being realised. In the models we are considering 
here, there is always a possible predator-prey interaction between any 
two species $i$ and $j$, either $a_{ij} > 0$ and $a_{ji} = 0$ or 
$a_{ji} > 0$ and $a_{ij} = 0$. Thus the ecosystems are potentially 
highly connected with one half of all inter-specific links realised. 
However, species usually focus on those prey to which they are 
best adapted. We therefore must introduce a mechanism
which would limit the number of species a predator actually
feeds on, with potential links becoming active only when the preferred 
prey is scarce or unavailable, or after a change in the composition of 
the web.  One way to (artificially) realise this feature is to allow 
only the best predators to feed on a given prey. We did this by replacing the
interactions $a_{ij}$ with adjusted interactions $a_{ij}'=a_j^{\rm
max}(1-(a_j^{\rm max}-a_{ij})/\delta)$ (Caldarelli {\it et al.}, 1998), 
with $\delta$ being a small parameter and $a_j^{\rm max}$ being the largest
interaction against $j$. Negative $a_{ij}'$ are set to zero. Fig.~\ref{fig2} 
shows a food web obtained from Lotka-Volterra dynamics with
this artificial constraint. Table 1 shows the mean number of species, of 
links per species, and the mean occupation numbers of the trophic levels 
for this model, for $R = 1\times10^4$, $\lambda = 0.1$, $c = 1.0$ and 
$\delta = 0.2$. For comparison, the results for the original Lotka-Volterra 
model with the parameters for $R = 1\times10^4$, $\lambda = 0.1$, 
and $c = 3.0$ are also shown. Similar web structures were obtained 
including direct inter-specific competition or using Holling and 
Beddington forms with the same type of adjusted interactions.


\begin{table}
\begin{center}
\begin{tabular}{|c||c|c|c|c|c|c|c|c|} \hline
 &  & & \multicolumn{6}{|c|} {Trophic level} \\ 
\cline{4-9}
Model & Number of & Links per & 1 & 2 & 3 & 4 & 5 & 6\\
\cline{4-9}
& species & species & \multicolumn{6}{|c|}{Number of species} \\
\cline{4-9}
& & & \multicolumn{6}{|c|}{Frequency of occupation} \\
\hline\hline
Lotka-Volterra without & 40.6(2.5) & 2.3(0.1) & 40.5 & 1.6 & --- & --- 
& --- & --- \\\cline{4-9}
adjusted interactions &           &          & 1.000 & 0.061 
& --- & --- & --- & ---  \\ 
\hline
Lotka-Volterra with & 69.8(20.4) & 1.4(0.1) & 20.9 & 22.1 & 19.1 & 7.5 & 1.4 & 
1.0 \\\cline{4-9}
adjusted interactions &            &          & 1.000 & 1.000 & 
1.000 & 0.987 & 0.140 & 0.003 \\ 
\hline
\hline
\end{tabular}
\end{center}
\caption[]{Food web statistics for the two Lotka-Volterra models: without 
and with adjusted interactions. The results are averaged over ten different 
simulations (lasting 200000 iterations) and over the last 20000 iterations 
of each simulation. The quantities in brackets give standard deviations over 
the ten runs for the number of species and links per species. Only links 
between non-environment species that constituted greater than 1\% of the 
predator's diet were counted in the calculation of the links per species.}
\end{table}



This shows that the ability of predators to concentrate on the prey
that they are best suited to exploit, rather than on all possible
prey, is essential for the production of realistic food web
structures. 

Finally, let us discuss evolutionary food web models with
ratio-dependent functional responses.  Ratio-dependent functional
responses naturally limit the number of actual prey of a predator by
dividing the available prey among its predators. One such
ratio-dependent form, given by (\ref{generalratio}), was suggested 
by Arditi and Michalski (1996). However, it leads to such
strong competition that only one predator species can feed on a given
prey species and thus cannot give rise to complex food web
structures. (One can conclude this directly from the published webs
(Arditi \& Michalski, 1996) or from an analytical calculation with two
predator species and one prey species). In reality, however, different
species exploit a food source in different ways and therefore
competition between individuals of different species is less than
between individuals of the same species.

This feature is implemented in the functional response (\ref{ourgij}),
which was introduced by some of us (Drossel {\it et al.}, 2001). 
Furthermore, equations (\ref{ourgij}) and (\ref{eff}) have
several other biologically meaningful features: when prey is very
abundant, the consumption rates saturate due to the term containing
$b$ in the denominator. Additionally, predators divide their effort
(or time) among the available prey in order to obtain the maximum
possible amount of food. The condition (\ref{eff}) is an
evolutionarily stable strategy and allows species to swiftly adjust to
a modified situation, thus endowing the food web model with the
flexibility found in real webs. Finally, equations (\ref{ourgij}) and
(\ref{eff}) are invariant if identical species are aggregated into a
single species. This feature is shared by the other models used in
this paper if the last term in equation (\ref{balance}) is either set
to zero or present for any pair of species but not if, for instance, 
only inter-specific competition is included.

Using the functional response (\ref{ourgij}) in the evolutionary
model, we rapidly obtain large and complex food webs with a stable
structure, although there is an ongoing turnover of species in the
system. The dynamics (\ref{balance}) converge quickly to a fixed
point.  The food webs generated by this model share many features with
real food webs, such as the fraction of top, intermediate and basal
species, and the mean number of links per species.  This is discussed
in detail elsewhere (Drossel {\it et al.}, 2001; Quince {\it et al.}, 2004a).  


\begin{figure}
\begin{center}
\includegraphics*[width=4.5cm]{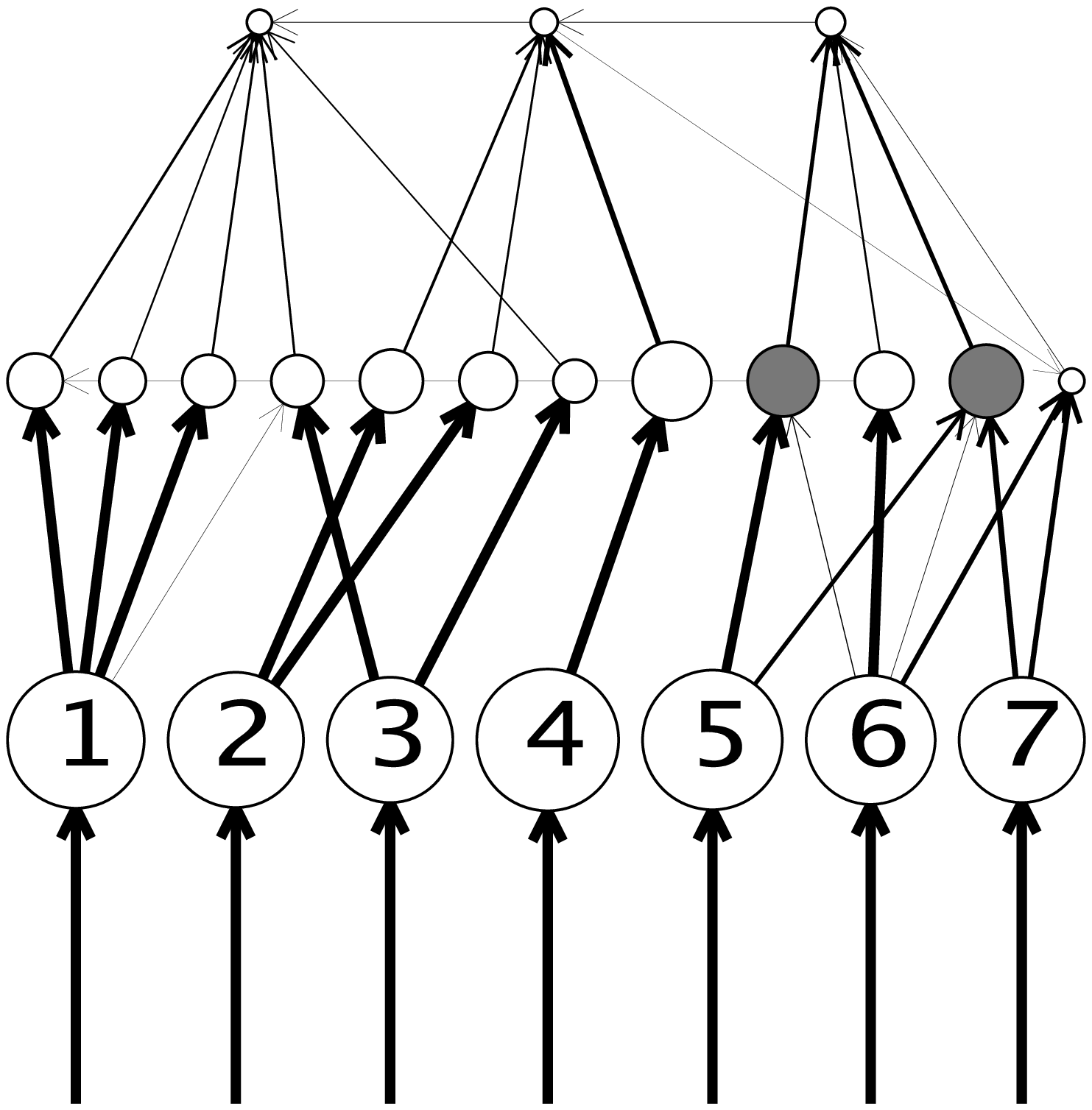} 
\hskip -5mm \includegraphics*[width=4.5cm]{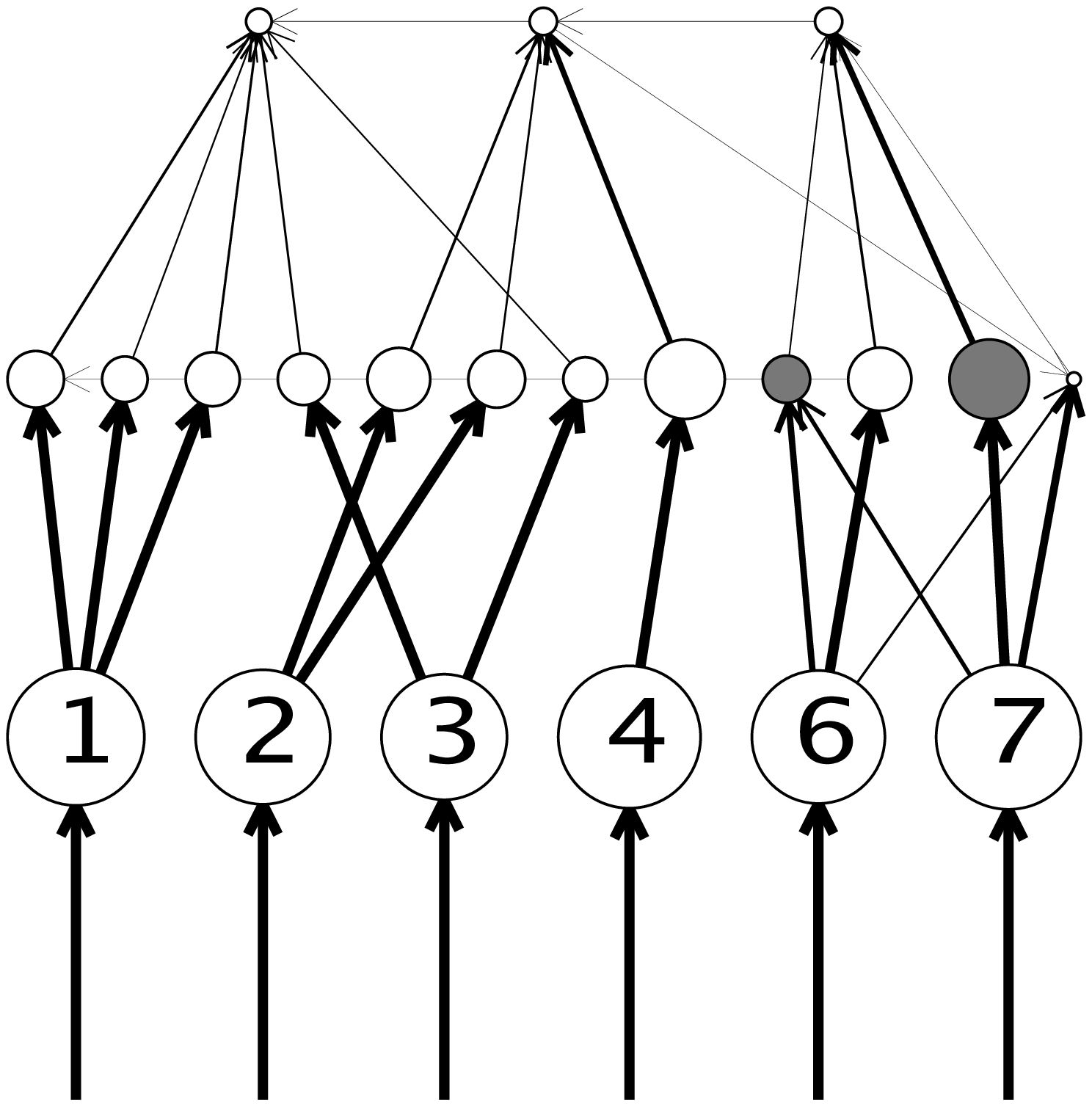}
\caption{A model web generated using the functional response (\ref{ourgij}) 
before (left) and after (right) deletion of basal species 5. 
The two shaded species feeding on it managed to survive by feeding at a 
higher rate on other species.
\label{fig3}}
\end{center}
\end{figure}


The webs generated by this model often  preserve their structure under 
changes in the species composition. Fig.~\ref{fig3} shows a web before and 
after a basal species was deleted. One can see that in this case all other 
species manage to survive, with the strength of several links being modified,
illustrating the flexibility and resilience of the model food webs. Of
course, the deletion of a randomly chosen species is not always
buffered in this perfect way. A useful measure of the ecosystem stability
to deletion is the fraction of species that can be removed without
causing further extinctions (Pimm, 1979). This statistic depends on 
the model parameters; for the values used to generate 
Fig.~\ref{fig3} it is 62\%. For the deletions which do cause further 
extinctions the distribution of event sizes decays roughly exponentially with
a characteristic size of just a few species (Quince {\it et al.}, 2004b).

\section{Simulation results: link strength distributions}

There has been an increasing realisation that food webs have a
large proportion of weak links (Paine, 1992; Tavares-Cromar \&
Williams, 1996; Berlow {\it et al.}, 1999; Neutel {\it et al.}, 2002), 
and that weak links tend to stabilise population dynamics 
(McCann {\it et al.}, 1998).  For this reason, we evaluated the link 
strength distribution in our model webs. We found that for all functional 
responses that generated large ecosystems either the link strength 
distributions were skewed towards zero or a large fraction of links were 
zero. This is a highly non-trivial result as in contrast to other work on 
the topic of weak links, the link strength distributions in our model 
are an emergent property of the system and not put in by hand. It is a 
strong indication that weak links are the natural outcome of long-term 
ecosystem evolution coupled to population dynamics. 

In the literature there are different definitions of link strength,
ranging from the biomass flowing through a link per unit time to the
response of one population size to a small change in another
population size (Laska \& Wootton, 1998). Here, we evaluated link
strength distributions based on the following two definitions: (i) the
proportion of the prey in the predator's diet after the population
dynamics had equilibrated,
\begin{equation} 
l_{ij}^{(1)}= \frac{g_{ij}}{\sum_k g_{ik}}\,,
\end{equation}
and (ii) the per-capita interaction strength $l_{ij}^{(2)}$. These are 
sometimes known as the elements of the ``community matrix'' (Laska \& Wootton, 
1998), although this phrase is also used to refer to the unnormalised matrix 
of partial derivatives (May 1973). It quantifies the strength of all direct 
interactions, both predator-prey interactions and interference competition. 
It is defined as:
\begin{equation}
l_{ij}^{(2)} = \frac{1}{N_{i}}\left(\frac{\partial \dot N_{i}}
{\partial N_{j}}\right).
\end{equation}
Definition (i) typically leads to a U-shaped link-strength distribution, 
with a peak at weak links and another one near the maximum link strength 1, 
indicating that many predators have one main prey. (Note that the second peak 
does not occur if the link strength is defined as the biomass passing through 
a link, since in this case the link strength is not normalised to 
$\sum_j l_{ij} = 1$). Definition (ii) typically leads to broad distributions 
with or without a peak at zero.

 
\begin{figure}
\begin{center}
\includegraphics*[width=9cm]{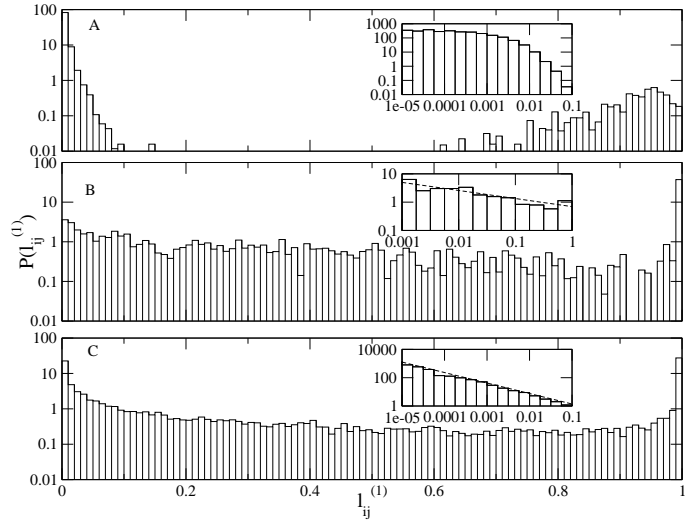}
\caption{The distribution of link strengths $l_{ij}^{(1)} > 0$ for the 
Lotka-Volterra model without adjusted scores (Fig.~A) with adjusted scores
(Fig.~B) and the ratio-dependent functional response (Fig.~C).
The $y-$axes on all graphs are scaled logarithmically. The insets show the 
data with both axes scaled logarithmically, for Figs.~B and C these include 
power-law fits which had exponents -0.74 and -0.29 respectively.
\label{fig4}}
\end{center}
\end{figure}


Fig.~\ref{fig4} shows the distribution of link strengths $l_{ij}^{(1)}$
for the two Lotka-Volterra models (without and with adjusted scores),
and for the model with the ratio-dependent functional response, averaged
over many evolutionary time steps and several different model
webs. For small link strength, the distribution for the
ratio-dependent functional response resembles a power law with an exponent
around $-0.74$, while the Beddington model with adjusted scores shows an 
exponent around $-0.9$, and the Lotka-Volterra model with adjusted scores 
shows an exponent around $-0.29$ (if a power law fit should be attempted at
all).  The Lotka-Volterra model without adjusted scores does not have several
trophic layers, and its link strength distribution appears to decay
exponentially at small values. For the Beddington functional response without 
adjusted scores, the decay is much steeper than with adjusted scores, but not 
exponential. All the above mentioned models are skewed towards
small values except for the Lotka-Volterra model with adjusted scores where
the approach to the origin is quite flat. For this model however only a small
proportion, ~3\%, of links have $l_{ij}^{(1)} > 0$. We conclude that
for the models capable of generating large ecosystems only a small fraction
of links were realised or $l_{ij}^{(1)}$ had a large weight at small values.
We can summarise this by calculating the fraction of links with 
$l_{ij}^{(1)} < 0.01$ including zero, which gives 91\% and 97\% for the 
Lotka-Volterra models (without and with adjusted scores) and 97\% for the 
ratio-dependent model.


\begin{figure}
\begin{center}
\includegraphics*[width=9cm]{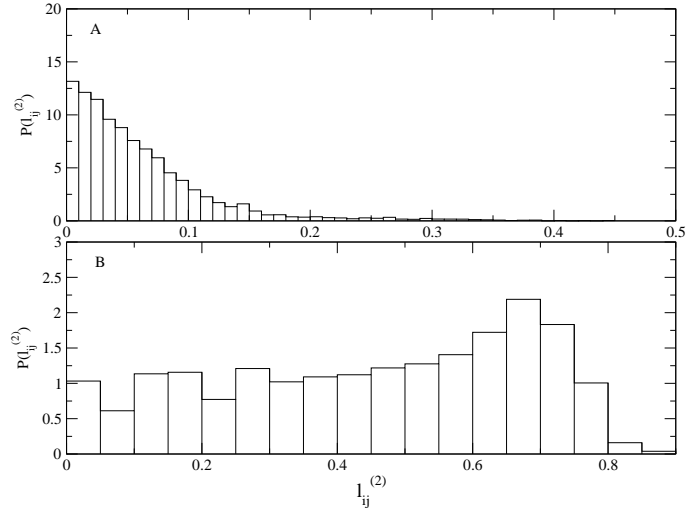}
\caption{The distribution of link strengths $l_{ij}^{(2)} > 0$ for the 
Lotka-Volterra models without adjusted scores (Fig.~A) and with 
adjusted scores (Fig.~B).
\label{fig5}}
\end{center}
\end{figure}


We can investigate these ideas further by examining the link strength 
according to the definition (ii); this can be either positive or negative. 
In the case of the Lotka-Volterra models, the only direct inter-specific 
interactions are between predators and prey with 
$l_{ij}^{(2)} = \lambda a_{ij}$ and $l_{ji}^{(2)} = - a_{ij}$ if $i$ 
predates $j$. Thus in Fig.~\ref{fig5} we only need to show the positive 
half of the $l_{ij}^{(2)}$ distribution. From this we see
that, whereas the Lotka-Volterra model without adjusted scores has a 
distribution of $l_{ij}^{(2)}$ values that is skewed towards zero,
the model with adjusted scores has a maximum at an intermediate value of 
$l_{ij}^{(2)}$. However, as was the case for $l_{ij}^{(1)}$, we
find a much higher proportion of non-zero links in the model without
adjusted interactions (51\% as compared to 3\%).


\begin{figure}[t]
\begin{center}
\includegraphics*[width=9cm]{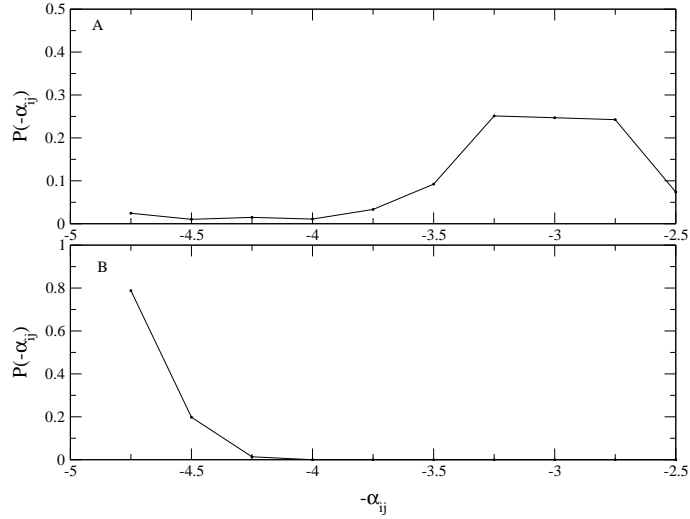}
\caption{The distribution of non-zero $-\alpha_{ij}$ for the Lotka-Volterra 
models with direct inter-specific competition. The results for the model 
without adjusted scores is shown in Fig.~A and with adjusted scores in Fig.~B.
The fraction of inter-specific interaction with non-zero $\alpha_{ij}$ was 
100\% for the former model and 23\% for the latter. Note that the non-zero
$\alpha_{ij}$ are clustered near to the smallest allowed magnitude of 2.5 
in Fig.~A and near to the largest allowed magnitude of 5 in Fig.~B.
\label{fig6}}
\end{center}
\end{figure}


If we introduce direct inter-specific competition into the Lotka-Volterra 
models as in Section 3 by setting $\alpha_{ij} = c\rho_{ij}(1 + q_{ij})/2$ 
in Eq.~(\ref{balance}), where $\rho_{ij}$ is equal to one if $i$ and $j$ share 
at least one prey and is zero otherwise and where $q_{ij}$ is the fraction of 
discrete features shared by the two species, we find that, even if adjusted 
interactions are used, only small webs can be evolved. The reason for this 
seems to be that the resulting direct competitive interactions which are 
allowed by the adjusted interactions are both strong and non-zero for a 
large fraction of species pairs in the web. This is illustrated by 
Fig.~\ref{fig6} where the distributions of the quantity $-\alpha_{ij}$ 
when $c = 5.0$ are plotted for the model with and without adjusted scores. 
We plot $-\alpha_{ij}$ because it is the contribution of direct competition 
to $l_{ij}^{(2)}$. Note that because $q_{ij}$ is discrete so is $\alpha_{ij}$
and that the range of non-zero $\alpha_{ij}$ is 
$-5 \leq -\alpha_{ij} \leq -2.5$. 


\begin{figure}[t]
\begin{center}
\includegraphics*[width=9cm]{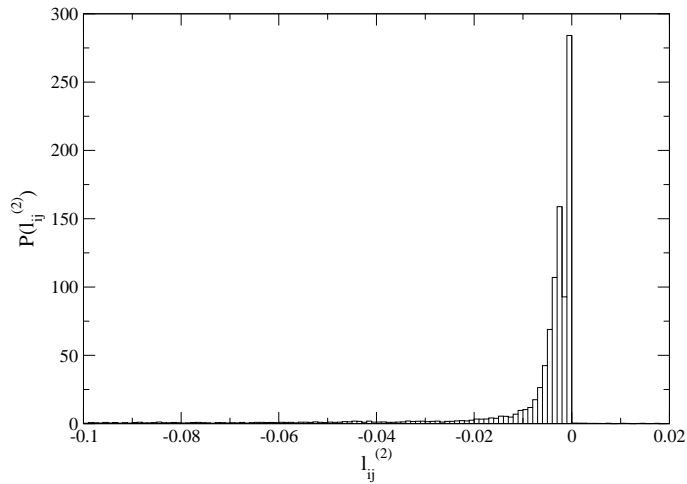}
\caption{The distribution of $l_{ij}^{(2)}$ for competitive interactions 
where the shared prey comprised greater than 1\% of both competitors diet 
estimated for the ratio-dependent model. This condition was satisfied by 
5\% of all species pairs.
\label{fig7}}
\end{center}
\end{figure}


The model with the ratio-dependent functional response, Eq.~(\ref{ourgij}),
also includes direct inter-specific competition, as a result of the implicit
sharing of prey between predators. However this does not lead to the strong 
competitive interactions seen in Fig.~\ref{fig6}, probably because 
this competition occurs through the predator-prey terms, which as was shown
in Fig.~\ref{fig4}, are themselves very diffuse. This can be seen in 
Fig.~\ref{fig7}, where the distribution of $l_{ij}^{(2)}$ values for all 
competitive interactions and where the shared prey comprised greater than 1\% 
of both competitors diet, is shown. This distribution, although bimodal, is 
heavily weighted towards the origin. This then explains why the ratio-dependent
model with inter-specific competition is, in contrast to the Lotka-Volterra 
models discussed immediately above, capable of evolving large complex 
ecosystems.

This finding of ours is complemented by a recent paper by Kondoh (2003), 
who investigates randomly linked food web models and the cascade model based 
on Lotka-Volterra dynamics with adjustable foraging efforts. He finds that 
incorporating adaptive foragers turns a negative relationship between 
stability and complexity, defined in terms of species number and potential 
connectance, into a positive one. This result had already been found, albeit 
over a more limited range of parameters, by Pelletier (2000). These results 
are strictly different to ours in that the stability criterion, community 
persistence defined as fraction of species surviving for some long time in 
a stochastic environment, is different. However the mechanism that allows 
the generation of large webs with high potential connectance, the reduction 
in the number of non-zero realised links and the reduction in strength of 
those links that are realised due to the foraging dynamics, is very similar 
to that proposed above. In fact this mechanism is arguably just May's 
original hypothesis, that strong links or a large fraction of links will 
destabilise ecosystems, placed in an adaptive context (May, 1972).

\section{Discussion}

The form that a realistic functional response might take has been the subject 
of a large number of papers in the literature but few, if any, of these 
suggestions were implemented in a model of multispecies communities in order 
to test their effectiveness. In this paper we have used an existing model 
to investigate the effect that different choices for the functional response 
have on food web structure. We studied models with a wide range of functional 
responses: Lotka-Volterra with and without direct inter-specific competition,
Holling and Beddington forms, all of these but with a mechanism to limit the 
number of species a predator actually feeds on, and ratio-dependent 
functional responses. We found that in the first two cases complex webs could 
not be built up, but in the last two cases they often could. More specifically,
unless an (artificial) mechanism was introduced which restricted predator
choice to prey which they were best suited to exploit or alternatively a 
ratio-dependent functional response was used, then stable webs consisting of
more than one trophic level could not be built up.  

Our second major conclusion was that, given we chose a functional response
which gave rise to complex, stable webs as described above, the link strength
distributions were skewed towards zero. For the Beddington functional response 
with adjusted scores or ratio-dependence, the link strength distribution
followed a power-law, but this was much less clear in the case of the 
Lotka-Volterra model, where only a small fraction of the links were realised
i.e. a large fraction of the links were zero. Some insight into the mechanisms 
involved in these effects was also gained by looking at the strength of a 
direct inter-specific competition introduced into the Lotka-Volterra equations.
Without adjusted scores, interactions were present, but not particularly 
strong. However with adjusted scores, about three-quarters of the interactions 
(for the parameter values we used) were zero, but the remaining quarter were 
strong. In this case we found that, even with adjusted scores, large complex
webs could not be grown. We concluded that the existence of strong competitive
interactions might also destabilise food webs.

In summary, we found that the type of functional response used in the 
population dynamics of multispecies communities has to be chosen carefully 
if a large complex community is to be sustained. If this is achieved, then 
a large proportion of weak links arises naturally from the evolution of the 
food webs. These conclusions were arrived at within the context of a class
of evolutionary food web models. It would be interesting to investigate if 
these results could be obtained from other starting points or with different 
model assumptions.

\vspace{0.9cm}

\noindent{\bf Acknowledgements}: We wish to thank Paul Higgs for useful
discussions. CQ thanks the EPSRC (UK) for financial support during the 
initial stages of this work.

\newpage

\section*{References}

\noindent Arditi, R., Michalski, J., 1996. Nonlinear food web
models and their responses to increased basal productivity. In
Food webs: Integration of patterns and dynamics. Polis, G.~A.,
Winemiller, K.~O. (eds), pp\,122--133. Chapman and Hall, New York.\hfill\break
\noindent Berlow, E.~L., Navarrete, S.~A., Briggs, C.~J., Power, M.~E.,
Menge, B.~A., 1999. Quantifying variation in the strengths of species
interactions. Ecology 80, 2206--2224.\hfill\break
\noindent Caldarelli, G., Higgs, P.~G., McKane, A.~J., 1998. Modelling 
coevolution in multispecies communities. J. Theor. Biol. 193, 345--358.
\hfill\break
\noindent Drossel, B., Higgs, P.~G., McKane, A.~J., 2001. The influence 
of predator-prey population dynamics on the long-term evolution of food 
web structure. J. Theor. Biol. 208, 91--107.\hfill\break
\noindent Kondoh, M.~K., 2003. Foraging adaptation and the relationship 
between food web complexity and stability. Science 299, 1388--1391.
\hfill\break
\noindent Laska, M.~S., Wootton, T.~J., 1998. Theoretical concepts and 
empirical approaches to measuring interaction strengths. Ecology 79, 
461--476. \hfill\break
\noindent L\"assig, M., Bastolla, U., Manrubia, S.~C., Valleriani, A., 2001.
Shape of ecological networks. Phys. Rev. Lett. 86, 4418--4421.
\hfill\break
\noindent Law, R., 1999. Theoretical aspects of community assembly. In:
Advanced ecological theory: principles and applications. McGlade, J. (ed).
pp\,143--171. Blackwell, Oxford.\hfill\break
\noindent May, R.~M., 1972. Will a large complex system be stable?
Nature 238, 413--414.\hfill\break
May, R.~M., 1973. Stability and complexity in model ecosystems. 
Princeton University Press, Princeton. \hfill\break
\noindent McCann, K., 2000. The diversity-stability debate. Nature
405, 228--233.\hfill\break
\noindent McCann, K, Hastings, A., Huxel, G.~R., 1998. Weak trophic 
interaction and the balance of nature. Nature 395, 794--798.
\hfill\break
\noindent Morton, R.~D., Law, R., 1997. Regional species pools and the 
assembly of local ecological communities. J. Theor. Biol. 187, 
321--331. \hfill\break
\noindent Neutel, A-M., Heesterbeek, J.~A.~P., de Ruiter, P.~C., 2002.
Stability in real food webs: weak links in long loops. Science
296, 1120--1123.\hfill\break
\noindent Paine, R.~T., 1992. Food-web analysis through field measurements of 
per capita interaction strength. Nature 355, 73--75.
\hfill\break
\noindent Pelletier, J.~D., 2000. Are large complex ecosystems more unstable? 
A theoretical reassessment with predator switching. Math. Biosc. 
163, 91--96. \hfill\break 
\noindent Pimm, S.~L., 1979. Complexity and stability: another look at 
MacArthur's original hypothesis. Oikos 33, 351--357.
\hfill\break
\noindent Pimm, S.~L., 1991. The balance of nature. The University of Chicago 
Press, Chicago. \hfill\break
\noindent Quince, C., Higgs, P.~G., McKane, A.~J., 2004a. Topological 
structure and interaction strengths in model food webs. Submitted for 
publication to Ecol. Model. \hfill\break
\noindent Quince, C., Higgs, P.~G., McKane, A.~J., 2004b. Deleting species 
from model food webs. Submitted for publication to Oikos. 
\hfill\break
\noindent Rosenzweig, M.~L., 1995. Species diversity in space and time. 
Cambridge University Press, Cambridge. \hfill\break
\noindent Tavares-Cromar, A.~F., Williams, D.~D., 1996. The importance of 
temporal resolution in food web analysis: evidence from a detritus-based 
stream. Ecol. Monogr. 66, 91--113.\hfill\break
\noindent Thompson, J.~N., 1998. Rapid evolution as an ecological process, 
TREE 13, 329--332. \hfill\break
\noindent Vandemeer, J. \emph{et al}., 2002. Increased competition may promote
species coexistence. Proc. Nat. Acad. Sci. 99, 8731--8736. \hfill\break
\noindent Yodzis, P., 1981. The stability of real ecosystems. Nature 
289, 674--676.

\end{document}